\documentclass[12pt]{article}

\usepackage{graphicx}
\usepackage{scicite}

\usepackage{times}
\newenvironment{sciabstract}{%
\begin{quote} \bf}
{\end{quote}}

\newcounter{lastnote}
\newenvironment{scilastnote}{%
\setcounter{lastnote}{\value{enumiv}}%
\addtocounter{lastnote}{+1}%
\begin{list}%
{\arabic{lastnote}.}
{\setlength{\leftmargin}{.22in}}
{\setlength{\labelsep}{.5em}}}
{\end{list}}

\title{Heat Capacity of a Strongly-Interacting Fermi Gas}

\author{J. Kinast, A. Turlapov, and J. E. Thomas$^{\ast}$\\
\\
\normalsize{Physics Department, Duke University, Durham, North
Carolina 27708-0305}\\
\\
\normalsize{$^\ast$To whom correspondence should be addressed;
E-mail:  jet@phy.duke.edu.}\\
\\
\normalsize{  }\\
\normalsize{(Submitted 10 September, 2004)}\\
\normalsize{}}
\date{}

\begin{document}


\baselineskip24pt


\maketitle


\begin{sciabstract}
We report on the measurement of the heat capacity for an
optically-trapped, strongly-interacting Fermi gas of atoms. In the
experiments, a precise input of energy to the gas is followed by
single-parameter thermometry. The thermometry determines a
temperature parameter $\tilde{T}$ from the best fit of a
Thomas-Fermi distribution with a  fixed Fermi radius  to the
spatial density of the cloud. At $\tilde{T}=0.33$, we observe a
transition between two patterns of behavior: For
$\tilde{T}=0.33-2.15$, we find that the heat capacity  closely
corresponds to that of a trapped normal Fermi gas of atoms with
increased mass. At low temperatures $\tilde{T}=0.04-0.33$, the
heat capacity clearly deviates from normal Fermi gas behavior.
\end{sciabstract}

Strongly-interacting, degenerate  atomic Fermi
gases~\cite{OHaraScience} provide a paradigm for strong
interactions in nature~\cite{AmScientist}. Measurements of the
interaction
energy~\cite{OHaraScience,MechStab,SalomonBEC,Grimmbeta} test
predictions of universal interactions in nuclear
matter~\cite{Heiselberg,Baker,Carlson}, as well as effective field
theories of strong interactions~\cite{Steele}. The anisotropic
expansion observed for strongly-interacting Fermi
gases~\cite{OHaraScience} is analogous to the ``elliptic flow" of
a quark-gluon plasma~\cite{Heinz}.  High temperature superfluidity
has been
predicted~\cite{Houbiers,CombescotHighTc,Holland,Timmermans,Griffin,Stajic,StrinatiTc}
in strongly-interacting Fermi gases, which can be used to test
theories of high temperature superconductivity~\cite{Levin}.
Microscopic evidence for high temperature superfluidity has been
obtained in the condensation of preformed
pairs~\cite{Jincondpairs,Ketterlecondpairs} and in radio frequency
measurements of the pairing gap~\cite{GrimmGap,JinGap}.
Macroscopic evidence  arises in anisotropic
expansion~\cite{OHaraScience} and in collective
excitations~\cite{Kinast,Bartenstein,KinastMagDep}.

In superconductivity and superfluidity, measurements of the heat
capacity~\cite{Keesom} have played  an exceptionally important
role in determining phase transitions~\cite{LondonBEC} and in
revealing the nature of the many-body quantum state of the system.
We report on the measurement of the heat capacity for a
strongly-interacting Fermi gas of $^6$Li atoms, confined in an
optical trap. Our experiments examine the fundamental
thermodynamics of the gas. In the following, we first describe how
the gas is prepared and our method for adding a precisely known
energy to the gas. Then we discuss our technique of thermometry,
which provides a monotonic temperature scale and a well-defined
method for comparing experiment with predictions.

We prepare a degenerate 50-50 mixture of the two lowest spin
states of $^6$Li atoms by forced evaporation in an ultrastable
CO$_2$ laser trap~\cite{OHaraStable} as described
previously~\cite{OHaraScience}.  At a bias magnetic field of 840
G, just above the Feshbach resonance, the trap depth is lowered by
a factor of $\simeq 580$ in a few
seconds~\cite{OHaraScience,Kinast} and then recompressed to 4.6\%
of the full trap depth in 1.0 s and held for 0.5 s to assure
equilibrium. After a controlled amount of energy is added to the
gas, as described below, the gas is allowed to thermalize for 0.1
s. Finally, the gas is released from the trap and imaged at 840 G
to determine the number of atoms and the temperature parameter
$\tilde{T}$. The column density is obtained by absorption imaging
of the expanded cloud after 1 ms time of flight, using a two-level
state-selective cycling transition~\cite{OHaraScience,Kinast}. In
the measurements, we take optical saturation into account exactly
and arrange to have very small optical pumping out of the
two-level system. For our trap, the total number of atoms is
$N=2.2(0.3)\times 10^5$. From the measured trap frequencies,
corrected for anharmonicity, we obtain
$\omega_\perp=\sqrt{\omega_x\omega_y} = 2\pi\times 1696(10)$ Hz
and $\omega_z=2\pi\times 72(5)$ Hz, so that
$\bar{\omega}=(\omega_x\omega_y\omega_z)^{1/3}=2\pi\times 592(14)$
Hz is the mean oscillation frequency. For these parameters, the
typical Fermi temperature $T_F=(3N)^{1/3}\hbar\bar{\omega}/k_B$
for a noninteracting gas is $\simeq 2.5\,\mu$K, small compared to
the final trap depth of $U_0/k_B=35\,\mu$K.

Energy is precisely added  to the trapped gas at fixed atom number
by releasing the cloud from the trap and permitting it to expand
for a short time $t_{heat}$ after which the gas is recaptured. As
shown below, even for the strongly-interacting gas, the energy
input is well-defined for very low initial temperatures, where
both the equation of state and the expansion dynamics are known.
During the times $t_{heat}$ used in the experiments, the axial
size of the gas changes negligibly, while transverse dimensions
expand by a factor $b_\perp(t_{heat})$. Hence, the harmonic
trapping potential energy in each of the two transverse directions
increases by a factor $b_\perp^2(t_{heat})$.

The initial potential energy is readily determined at zero
temperature. This follows  from the equation of state of the gas,
$(1+\beta)\epsilon_F(\mathbf{x})+U_{trap}(\mathbf{x})=\mu_0$~\cite{OHaraScience,MechStab,eqofstate},
where $\epsilon_F(\mathbf{x})$ is the local Fermi energy, $\beta$
is the unitary gas
parameter~\cite{OHaraScience,Heiselberg,MechStab,Carlson,Strinati},
$U_{trap}$ is the harmonic approximation to the trapping
potential, and $\mu_0$ is the global chemical potential. The
equation of state is equivalent to that of a harmonically trapped
noninteracting gas of particles with an effective
mass~\cite{Baker}, which in our notation is $M^*=M/(1+\beta)$,
where $M$ is the bare mass. Since the gas behaves as a harmonic
oscillator, the mean potential energy is half of the total energy.
As $\beta <0$~\cite{Carlson}, $M^*>M$, so that the effective
oscillation frequencies and the chemical potential are simply
scaled down, i.e.,
$\mu_0=k_BT_F\sqrt{1+\beta}$~\cite{OHaraScience,MechStab}. The
total energy at zero temperature, which determines the energy
scale,  is therefore
\begin{equation}
E_0=\frac{3}{4}N\mu_0=\frac{3}{4}Nk_BT_F\sqrt{1+\beta}.
\label{eq:E0}
 \end{equation}
For each direction, the initial potential energy at zero
temperature is  $E_0/6$. Then, the total energy of the gas after
heating is given by~\cite{anharmonicpotential},
\begin{equation}
E(t_{heat})=\eta\,E_0\left[\,\frac{2}{3}+\frac{1}{3}\,b_\perp^2(t_{heat})\right].
\label{eq:energy}
\end{equation}
Here, $\eta$ is a correction factor arising from the finite
temperature of the gas prior to the energy input. For the
noninteracting gas, $\eta_{nonint}$ is determined at the lowest
temperature $\tilde{T}=0.23$ from the energy for an ideal Fermi
gas. For the strongly-interacting gas, where the initial
temperature is very low and  $\tilde{T}=0.04$, we assume a
Sommerfeld correction~\cite{Ashcroft}  and obtain $\eta_{int}
\simeq  1+ 2\pi^2 \tilde{T}^2/3\simeq 1.01$, which hardly affects
the energy scale.

 The strongly-interacting gas exhibits
hydrodynamic, anisotropic expansion~\cite{OHaraScience}, so that
$b_\perp=b^H_\perp$ is a hydrodynamic expansion
factor~\cite{OHaraScience,Menotti}. For the noninteracting gas, we
use a ballistic expansion factor
$b^B_\perp(t)=\sqrt{1+(\omega_\perp t)^2}$.  The temperature
change during the expansion time $t_{heat}\leq 460\,\mu$s must be
very small, since the minimum value of $\tilde{T}=0.04$ is
measured by imaging the interacting cloud after 1 ms of expansion.
Hence, the primary heating arises only after recapture and
subsequent equilibration.

Thermometry of strongly interacting Fermi gases is not well
understood. By contrast, thermometry of noninteracting Fermi gases
can be simply accomplished by fitting the spatial distribution of
the cloud with a Thomas-Fermi (T-F) profile, which is a function
of two parameters. We choose them to be the Fermi radius
$\sigma_x$ and the reduced temperature $T/T_F$. However, this
method is only precise at temperatures well below $0.5\, T_F$,
where $\sigma_x$ and $T/T_F$ are determined independently. At
higher temperatures, where the Maxwell-Boltzmann limit is
approached, such a fit determines only the product $\sigma_x^2\,
T/T_F$. We  circumvent this problem by determining $\sigma_x$ from
a low temperature fit, and then hold it constant in the fits at
all higher temperatures, enabling a one-parameter determination of
the reduced temperature.

For strongly interacting Fermi gases below the superfluid
transition temperature $T_c$, the spatial profile may contain
normal and superfluid components~\cite{Stajic}. However,
experimentally and theoretically, one finds that the spatial
profile of a strongly interacting gas closely resembles a T-F
distribution~\cite{OHaraScience,LevinDensity}. For this reason,
T-F fits to the cloud profiles are commonly used to estimate the
reduced temperature, which is often reported as $T/T_F$, where
$T_F$ is the Fermi temperature for a noninteracting gas. Analogous
to the noninteracting case, we define an experimental
dimensionless temperature parameter $\tilde{T}$, which is to be
determined by fitting the cloud profiles with a T-F
distribution~\cite{Jackson}, holding constant the Fermi radius of
the interacting gas, $\sigma_x'$.
 Unlike  two parameter fitting procedures, this single parameter
method is stable. We find experimentally that $\tilde{T}$
increases monotonically from the highly degenerate regime to the
Maxwell-Boltzmann limit. This fitting procedure also leads us to
define a natural reduced temperature scale
\begin{equation}
\tilde{T}_{nat}\equiv\frac{k_BT}{\mu_0}=\frac{T}{T_F\sqrt{1+\beta}},
\label{eq:temp}
\end{equation}
which is consistent with our choice of fixed Fermi radius
$\sigma_x'$, i.e., $M\omega_x^2\sigma_x'^2/2=\mu_0$. At high
temperatures, we must interpret $\tilde{T}=\tilde{T}_{nat}$, to
obtain the correct Maxwell-Boltzmann limit. At low temperatures,
$\tilde{T}\simeq \tilde{T}_{nat}$ yields an estimate of $T/T_F$.
However, to determine the precise correspondence between
$\tilde{T}$ and the reduced temperature $T/T_F$ which is input
into theoretical models,  one should perform the experimental
fitting procedure with the theoretically generated density
profiles as suggested and implemented  by Chen et
al.,~\cite{ChenScience}.

The experimental fitting procedure measures $\tilde{T}$ by first
obtaining one dimensional, transverse spatial distributions $n(x)$
from the column density by spatially integrating along the trap
axial direction. Dividing by the total number of atoms per spin
state, we obtain normalized spatial profiles. Then $\tilde{T}$ is
determined using the one parameter T-F fit method, yielding
0.04--2.15 for the strongly-interacting gas and 0.2--1.1 for the
noninteracting gas.

The experimental energy scale Eq.~\ref{eq:energy} and the natural
temperature scale Eq.~\ref{eq:temp} are determined by calculating
$\beta$ from the measured Fermi radii for the interacting and
noninteracting gas samples. The relation is given by
$\sigma'_x=\sigma_x(1+\beta)^{1/4}$~\cite{MechStab}, where
$\sigma_x=\sqrt{2k_BT_F/(M\omega_x^2)}$ is the radius for a
noninteracting gas. To determine  $\sigma_x'$, we measure the size
of the cloud after 1 ms of expansion, and scale it down by the
known hydrodynamic expansion factor of
$b_H(1\,\mbox{ms})=13.3$~\cite{OHaraScience,Menotti}. We then
determine the Fermi radius
$\sigma_x'=11.98\,(N/2)^{1/6}\,\mu\mbox{m}/13.3=0.901(0.021)\,(N/2)^{1/6}\mu$m.
 Using $\sigma_x=1.065\, (N/2)^{1/6}\,\mu$m for our
trap parameters,  yields $\beta =-0.49(0.04)$~\cite{error} in
reasonable agreement with the best current predictions, where
$\beta=-0.56$~\cite{Carlson}, and $\beta =-0.545$~\cite{Strinati}.

We now apply our energy input and thermometry methods to measure
the heat capacity of an optically trapped Fermi gas, i.e., for
different values of $t_{heat}$, we measure the temperature
parameter $\tilde{T}$  and calculate the total energy
$E(t_{heat})/E_0$ from Eq.~\ref{eq:energy}. To obtain high
resolution data, 30-40 different heating times $t_{heat}$ are
chosen. The data for each of these heating times are acquired in a
random order to minimize systematic error. Ten complete runs are
taken through the entire random sequence.

\begin{figure}
\begin{center}\
\includegraphics[bb= 105 299 475 496]{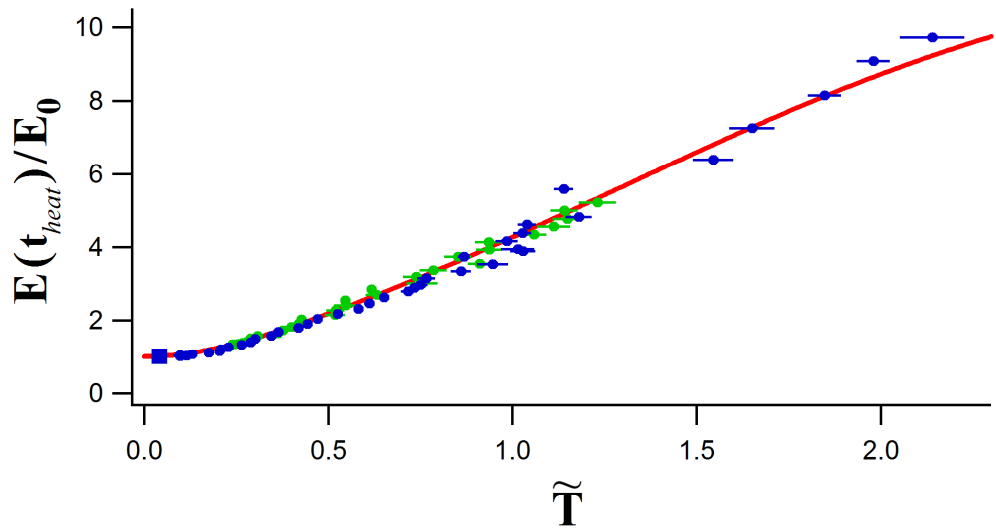}
\end{center}
\caption{Total energy  versus temperature. For each heating time
$t_{heat}$, the temperature parameter $\tilde{T}$ is measured from
the cloud profile, and the total energy $E(t_{heat})$ is
calculated from Eq.~\ref{eq:energy} in units of the ground state
energy $E_0$. Green circles: noninteracting Fermi gas data;
 Blue circles: strongly interacting Fermi gas data. Red curve: predicted
 energy versus reduced temperature for a noninteracting, trapped
 Fermi gas, $E_{ideal}(\tilde{T})/E_{ideal}(0)$.
\label{fig:LinTemp}}
\end{figure}

To test the method with a known system, we first measure the heat
capacity for a noninteracting Fermi gas at 526 G. The gas is
initially cooled to $\tilde{T}=0.23$ (the lowest temperature we
can achieve in this case) by 30 seconds of forced evaporation at
300 G as described previously~\cite{Kinast}, and then heated as
described above. Fig.~\ref{fig:LinTemp} shows the data (green
dots) which represent the calculated $E(t_{heat})/E_0$ versus the
measured value of $\tilde{T}$, for each $t_{heat}$. For
comparison, predictions for a noninteracting, trapped Fermi gas,
$E_{ideal}(\tilde{T})/E_{ideal}(0)$ are shown as the red curve,
where $\tilde{T}=T/T_F$ in this case. Here, the chemical potential
and energy are calculated using a finite temperature Fermi
distribution and the density of states for the trapped gas. We use
the density of states for a gaussian potential
well~\cite{OHaraStable}, rather than the harmonic oscillator
approximation. This yields  very good agreement at all
temperatures.

Next, we measure the heat capacity for the strongly interacting
gas at 840 G. Here the gas is cooled to $\tilde{T}=0.04$ and then
heated. Fig.~\ref{fig:LinTemp} shows the data (blue dots) which
represent $E(t_{heat})/E_0$ versus the measured value of
$\tilde{T}$, for each $t_{heat}$. Note that the temperature
parameter $\tilde{T}$ varies by a factor of 50 and the total
energy by a factor of 10. Remarkably, on a large scale plot, the
data for the strongly interacting and noninteracting gases appear
quite similar.

\begin{figure}
\begin{center}\
\includegraphics[bb= 91 224 486 560]{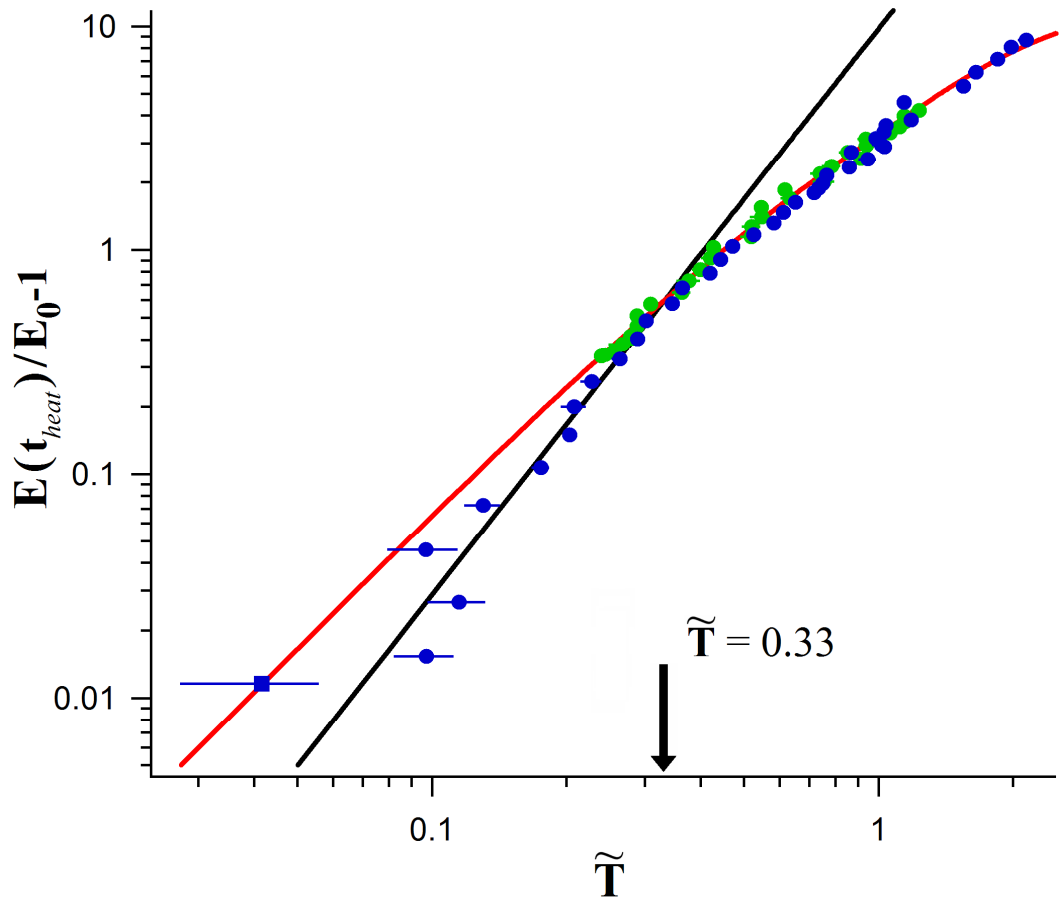}
\end{center}
\caption{Energy input versus temperature from
Fig.~\ref{fig:LinTemp} on a $log-log$ scale. The strongly
interacting Fermi gas shows a transition in behavior near
$\tilde{T}=0.33$. Green circles: noninteracting Fermi gas data;
Blue circles: strongly interacting Fermi gas data. Red curve,
prediction for a noninteracting, trapped Fermi gas. Black line,
best fit power law $9.8\,\tilde{T}^{2.53}$. Note the lowest
temperature point (blue square) is not included in the fits, as it
is constrained to lie on the red curve.\label{fig:LogTemp}}
\end{figure}

A striking result is shown by plotting the data for the strongly
interacting gas on a $log-log$ scale. Fig.~\ref{fig:LogTemp} shows
that the data reveal a transition in behavior at $\tilde{T}\simeq
0.33$, where the slope changes. Above $\tilde{T}\simeq 0.33$, the
data for strongly interacting data overlap closely with that of
the noninteracting gas. Below $\tilde{T}\simeq 0.33$, the data
deviates significantly from noninteracting Fermi gas behavior.
This transition may arise from changes in the behavior of the
total energy and from changes in the spatial profile of the gas
which serves as our thermometer.

Insights into the microscopic structure of the strongly
interacting gas can be obtained from the temperature scaling of
the energy. Above the transition, for $\tilde{T}\geq 0.33$, we
find that the data in
Figures~\ref{fig:LinTemp}~and~\ref{fig:LogTemp} are well fit by
$E(\tilde{T})=\sqrt{1+\beta}\,E_{ideal}(\tilde{T})$, with a
constant $\beta=-0.49$. This suggests that
$\tilde{T}=\tilde{T}_{nat}\equiv T/(T_F\sqrt{1+\beta})$ is a good
approximation above the transition. Such scaling may be a
manifestation of universal
thermodynamics~\cite{HoUniversalThermo}.

Below the transition, for $\tilde{T}\leq 0.33$, the gas may
comprise several components, for example, a normal Fermi gas and
both superfluid and noncondensed pairs, each contributing
differently to the temperature scaling, as arises in a pseudogap
model~\cite{Stajic,ChenScience}. For simplicitiy,  we consider
here a temperature scaling of the form $E(\tilde{T})/E_0=1+
b\,\tilde{T}^c$. For sufficiently low temperature, one expects
$c=2$ for an ideal Fermi gas, $c=5/2$ for a homogeneous
noninteracting Bose gas, and $c=4$ for a harmonically trapped Bose
gas. The best fit (black line in Fig.~\ref{fig:LogTemp})
corresponds to $c=2.53(0.15)$ and $b = 9.8(1.9)$. The $\chi^2$ per
degree of freedom for this fit is 1.4. We find that the parameters
$b$ and $c$ are strongly correlated. Holding $c=5/2$, we obtain
$b= 9.4(0.2)$. Fitting a quadratic temperature dependence yields
$b=4.8 (0.2)$, and a larger $\chi^2$ per degree of freedom of 5.2.
A $T^4$ power law fit yields $b=63.8 (4.1)$ and a $\chi^2$ per
degree of freedom of 7.1. These results suggest that the gas is
neither a normal Fermi gas nor a BEC of small weakly interacting
molecules.

One can understand 5/2 power scaling at very low temperature as
arising from thermal excitation of low energy bosons (fermion
pairs)~\cite{Levin}, where the fermionic contribution is
exponentially suppressed by the superfluid gap. A simple picture
of the $5/2$ power scaling is that short wavelength thermal
excitations increase the local kinetic energy of bound pairs
without breaking them, yielding the density of states and energy
for free particles in three dimensions. To make an estimate of $b$
based on universal scaling, we assume that the bosons have a mass
of $2M^*$, so that the density of states per unit volume for a
locally homogeneous gas is
$(4M^*/\hbar^2)^{3/2}\epsilon^{1/2}/(4\pi^2)$. The total energy is
easily determined using a Bose distribution with zero chemical
potential. Assuming that the pairing energy scale is large
compared to $k_BT$ over most of the trap volume, the $\epsilon$
integration is approximated from 0 to $\infty$. Multiplying the
resulting energy density by the trap volume $N/\bar{n}$, where
$\bar{n}$ is the average density, we obtain
$E/E_0=1+b\,\tilde{T}_{nat}^{5/2}$, where
$b=(3/4)\zeta(5/2)(2\pi)^{3/2}(315/512)=9.75$, close to the result
$9.4 (0.2)$ obtained from the fit.

 We estimate the transition temperature  from the intersection
 point, $\tilde{T}=0.33(.02)$~\cite{error}, of the power law fit and the scaled ideal gas prediction, Fig.~\ref{fig:LogTemp}.
 To extract a preliminary  value of $T_c/T_F$, we assume $\tilde{T}=\tilde{T}_{nat}$ near the transition temperature, and use the measured value of $\beta
 =-0.49$. We then obtain
 $T_c/T_F=\sqrt{1+\beta}\,\tilde{T}=0.24(.02)$~\cite{error}, close to predictions for the superfluid transition temperature which have
 been made over the last decade~\cite{Levin,Randeria,StrinatiTc,Torma}. The
 fractional  change in the heat capacity $C$ is estimated from the slope change in the fits to the data, assuming
 that the temperature calibration function is smooth near $T_c$~\cite{ChenScience}. In that
 case, $(C_>-C_<)/C_>=-0.48(0.03)$, where $>(<)$ denotes above (below) $T_c$.

Recently, Q. Chen, J. Stajic  and K. Levin have done a pseudogap
model of a trapped, strongly interacting Fermi
gas~\cite{ChenScience}, and obtain both the energy  and the
spatial profile  as a function of reduced temperature $T/T_F$,
throughout the superfluid and normal region. The temperature scale
$T/T_F$ is calibrated to our $\tilde{T}$ by fitting one
dimensional T-F profiles to the theoretical spatial distributions
as described above, yielding a monotonic relation. The data  of
Figures~\ref{fig:LinTemp}~and~\ref{fig:LogTemp} are very well
reproduced by the theory.


\begin{thebibliography}{10}

\bibitem{OHaraScience}
K.~M. O'Hara, S.~L. Hemmer, M.~E. Gehm, S.~R. Granade, J.~E.
Thomas, {\it
  Science\/} {\bf 298}, 2179 (2002).

\bibitem{AmScientist}
M.~E. Gehm, J.~E. Thomas, {\it Am. Scientist\/} {\bf 92}, 238
(2004).

\bibitem{MechStab}
M.~E. Gehm, S.~L. Hemmer, S.~R. Granade, K.~M. O'Hara, J.~E.
Thomas, {\it Phys.
  Rev. A\/} {\bf 68}, 011401(R) (2003).

\bibitem{SalomonBEC}
T.~Bourdel, {\it et~al.\/}, {\it Phys. Rev. Lett.\/} {\bf 93},
050401 (2004).

\bibitem{Grimmbeta}
M.~Bartenstein, {\it et~al.\/}, {\it Phys. Rev. Lett.\/} {\bf 92},
120401
  (2004).

\bibitem{Heiselberg}
H.~Heiselberg, {\it Phys. Rev. A\/} {\bf 63}, 043606 (2001).

\bibitem{Baker}
J.~G.~A.~Baker, {\it Phys. Rev. C\/} {\bf 60}, 054311 (1999).

\bibitem{Carlson}
J.~Carlson, S.-Y. Chang, V.~R. Pandharipande, K.~E. Schmidt, {\it
Phys. Rev.
  Lett.\/} {\bf 91}, 050401 (2003).

\bibitem{Steele}
J.~V. Steele, Effective field theory power counting at finite
density (2000).
  Nucl-th/0010066.

\bibitem{Heinz}
P.~F. Kolb, U.~Heinz, {\it Quark Gluon Plasma 3\/} (World
Scientific, 2003), p.
  634. See Hydrodynamic Description of Ultrarelativistic Heavy Ion Collisions,
  arXiv: nucl-th/0305084.

\bibitem{Houbiers}
M.~Houbiers, {\it et~al.\/}, {\it Phys. Rev. A\/} {\bf 56}, 4864
(1997).

\bibitem{CombescotHighTc}
R.~Combescot, {\it Phys. Rev. Lett\/} {\bf 83}, 3766 (1999).

\bibitem{Holland}
M.~Holland, S.~J. J. M.~F. Kokkelmans, M.~L. Chiofalo, R.~Walser,
{\it Phys.
  Rev. Lett.\/} {\bf 87}, 120406 (2001).

\bibitem{Timmermans}
E.~Timmermans, K.~Furuya, P.~W. Milonni, A.~K. Kerman, {\it Phys.
Lett. A\/}
  {\bf 285}, 228 (2001).

\bibitem{Griffin}
Y.~Ohashi, A.~Griffin, {\it Phys. Rev. Lett.\/} {\bf 89}, 130402
(2002).

\bibitem{Stajic}
J.~Stajic, {\it et~al.\/}, {\it Phys. Rev. A\/} {\bf 69}, 063610
(2004).

\bibitem{StrinatiTc}
A.~Perali, P.~Pieri, L.~Pisani, G.~C. Strinati, {\it Phys. Rev.
Lett.\/} {\bf
  92}, 220404 (2004).

\bibitem{Levin}
Q.~Chen, J.~Stajic, S.~Tan, K.~Levin, {BCS}-{BEC} crossover: From
high
  temperature superconductors to ultracold superfluids (2004).
  ArXiv:cond-mat/0404274.

\bibitem{Jincondpairs}
C.~A. Regal, M.~Greiner, D.~S. Jin, {\it Phys. Rev. Lett.\/} {\bf
92}, 040403
  (2004).

\bibitem{Ketterlecondpairs}
M.~W. Zwierlein, {\it et~al.\/}, {\it Phys. Rev. Lett.\/} {\bf
92}, 120403
  (2004).

\bibitem{GrimmGap}
C.~Chin, {\it et~al.\/}, {\it Science\/} {\bf 305}, 1128 (2004).

\bibitem{JinGap}
M.~Greiner, C.~A. Regal, D.~S. Jin, Probing the excitation
spectrum of a fermi
  gas in the bcs-bec crossover regime (2004). ArXiv:cond-mat/0407381.

\bibitem{Kinast}
J.~Kinast, S.~L. Hemmer, M.~E. Gehm, A.~Turlapov, J.~E. Thomas,
{\it Phys. Rev.
  Lett.\/} {\bf 92}, 150402 (2004).

\bibitem{Bartenstein}
M.~Bartenstein, {\it et~al.\/}, {\it Phys. Rev. Lett.\/} {\bf 92},
203201
  (2004).

\bibitem{KinastMagDep}
J.~Kinast, A.~Turlapov, J.~E. Thomas, Breakdown of hydrodynamics
in the radial
  breathing mode of a strongly-interacting fermi gas (2004).
  ArXiv:cond-mat/0408634; To appear in Phys. Rev. A Rapid Comm. {\bf 70}.

\bibitem{Keesom}
W.~H. Keesom, K.~Clusius, {\it Proc. Roy. Acad. (Amsterdam)\/}
{\bf 35}, 307
  (1932).

\bibitem{LondonBEC}
F.~London, {\it Phys. Rev.\/} {\bf 54}, 947 (1938).

\bibitem{OHaraStable}
K.~M. O'Hara, {\it et~al.\/}, {\it Phys. Rev. Lett.\/} {\bf 82},
4204 (1999).

\bibitem{eqofstate}
This zero temperature equation of state for a unitary gas is
supported by the
  observed spatial profiles at low temperatures, which are well approximated by
  a zero temperature T-F distribution as observed~\cite{OHaraScience} and
  predicted~\cite{Carlson,Strinati,Levin}. Further support arises from the
  measured radial breathing mode frequency at 840 G near
  resonance~\cite{KinastMagDep}, which is in excellent agreement with
  predictions~\cite{Stringariosc,Heiselbergosc} for a unitary, hydrodynamic
  Fermi gas.

\bibitem{Strinati}
A.~Perali, P.~Pieri, G.~C. Strinati, {\it Phys. Rev. Lett.\/} {\bf
93}, 100404
  (2004).

\bibitem{anharmonicpotential}
After the cloud expands for a time $t_{heat}$, the energy changes
when the
  trapping potential $U(\mathbf{x})$ is abruptly restored, i.e., $\Delta
  E(t_{heat})=\int
  d^3\mathbf{x}[n(\mathbf{x},t_{heat})-n_0(\mathbf{x})]U(\mathbf{x})$, where
  $n(\mathbf{x},t_{heat})$ is the density of the expanded cloud, which is
  related by a scale transformation~\cite{OHaraScience,Menotti} to the density
  prior to release, which takes the form of a zero temperature T-F profile
  $n_0(x,y,z)$~\cite{eqofstate}. Using this method, we obtain
  Eq.~\ref{eq:energy} as well as the anharmonic correction $\Delta E$ arising
  for a gaussian beam trapping potential. For a cylindrically symmetric trap:
  $\Delta
  E/E_0=-\mu_0[2b_\perp^4(t)+b_\perp^2(t)-3]/(30U_0)+\mu_0^2[4b_\perp^6(t)+2
  b_\perp^4(t)+3 b_\perp^2(t)-9]/(360 U_0^2)$. Note that for our experiments,
  we assume a gaussian beam potential with three different dimensions.

\bibitem{Ashcroft}
N.~W. Ashcroft, N.~D. Mermin, {\it Solid State Physics\/} (Holt,
Rinehart, and
  Winston, New York, 1976).

\bibitem{Menotti}
C.~Menotti, P.~Pedri, S.~Stringari, {\it Phys. Rev. Lett.\/} {\bf
89}, 250402
  (2002).

\bibitem{LevinDensity}
J.~Stajic, Q.~Chen, K.~Levin, Measuring condensates in fermionic
superfluids
  via density profiles in traps (2004). ArXiv:cond-mat/0408104.

\bibitem{Jackson}
B.~Jackson, P.~Pedri, S.~Stringari, {\it Europhys. Lett.\/} {\bf
67}, 524
  (2004). Our fit method is derived from ideas presented in this paper.

\bibitem{ChenScience}
Q.~Chen, J.~Stajic, K.~Levin, Thermodynamics of ultracold fermions
in traps
  (2004). ArXiv:cond-mat/0411090.

\bibitem{error}
The quoted errors in this paper are statistical only and represent
one standard
  error.

\bibitem{HoUniversalThermo}
T.-L. Ho, {\it Phys. Rev. Lett.\/} {\bf 92}, 090402 (2004).

\bibitem{Randeria}
C.~A. R.~S. de~Melo, M.~Randeria, J.~R. Engelbrecht, {\it Phys.
Rev. Lett.\/}
  {\bf 71}, 3202 (1993).

\bibitem{Torma}
J.~Kinnunen, M.~Rodr\mbox{\'{i}}guez, P.~T\mbox{\"{o}rm\"{a}},
{\it Science\/}
  {\bf 305}, 1131 (2004).

\bibitem{Stringariosc}
S.~Stringari, {\it Europhys. Lett.\/} {\bf 65}, 749 (2004).

\bibitem{Heiselbergosc}
H.~Heiselberg, {\it Phys. Rev. Lett.\/} {\bf 93}, 040402 (2004).

\end{thebibliography}


\begin{scilastnote}
\item We thank K. Levin, Q. Chen, and T.-L. Ho for stimulating
correspondence, and for providing physical insights on the
temperature dependence of the energy.  This research is supported
by the Chemical Sciences, Geosciences and Biosciences Division of
the Office of Basic Energy Sciences, Office of Science, U. S.
Department of Energy, the Physics Divisions of the Army Research
Office  and the National Science Foundation, and the Fundamental
Physics in Microgravity Research program of the National
Aeronautics and Space Administration.
\end{scilastnote}

\clearpage

\end{document}